\documentstyle{article}
\begin{document}

\title{Is the broad histogram random walk dynamics correct?}

\author{Jian-Sheng Wang\\
Department of Computational Science,\\
National University of Singapore,\\
Singapore 119260}

\date{2 October 1998}

\maketitle

\begin{abstract}
We show explicitly that the broad histogram single-spin-flip
random walk dynamics does not give correct microcanonical
average even in one dimension.  The dynamics violates detailed
balance condition by an amount which decreases with system size.
As a result, in distribution different configurations with the
same energy can have different probabilities.  We propose a
modified dynamics which ensures detailed balance and the
histogram obtained from this dynamics is exactly flat.
\end{abstract}

Two years ago, Oliveira et al
\cite{Oliveira,Oliveira-int,Oliveira-exact,Oliveira-ccp}
proposed an interesting method for Monte Carlo simulation of
statistical mechanical systems.  The number of degenerate
states $n(E)$ is computed from equations relating $n(E)$ to
the number of type of potential moves of a given dynamics.
The Monte Carlo computation is purely combinatorial (or
geometrical) unrelated to thermodynamics.  The temperature
dependence enters only after simulation in the weighting
formulas.  It is indeed an efficient method in comparison with
histogram \cite{Salzburg,Ferrenberg-Swendsen}, especially
multiple histogram method \cite{Ferrenberg-Swendsen}.

In the following, we use the Ising model to illustrate the
method, but the idea generalizes easily to other models.  One
first chooses a protocol of Monte Carlo moves, for example,
the single-spin flips.  For a given state $\sigma$ with energy
$E=E(\sigma)$ ($\sigma$ without an index denotes the set of
all spins), consider all possible single-spin flips.  The
single-spin flips change the current state into $N$ possible
new states, with new energy $E'=E(\sigma')= E +
\Delta E$ taking a small set of values. 
We classify the $N$ new states according to $\Delta E$, and
count the number of such moves $N(\sigma, \Delta E)$.  We
have $\sum_{\Delta E} N(\sigma, \Delta E) = N$.  Then one can
show rigorously that \cite{Oliveira-exact,Berg-Hansmann,Wang}
\begin{equation}
n(E) \bigl\langle N(\sigma, \Delta E)\bigr\rangle_E 
  = n(E+\Delta E) \bigl\langle 
N(\sigma', -\Delta E)\bigr\rangle_{E + \Delta E}
\label{EqBH}
\end{equation}
as long as a set of moves and their reverse moves are both
allowed.  The average is a microcanonical average, i.e.,
average over all configurations having a fixed energy with
equal weight:
\begin{equation}
\bigl\langle A(\sigma) \bigr\rangle_E = { 1 \over n(E) } 
  \sum_{E(\sigma) = E} A(\sigma).
\end{equation}
Equation (\ref{EqBH}) is the fundamental result of broad
histogram method.  Once the quantity $\langle N(\sigma, \Delta
E) \rangle_E$ is computed by any means, the density of states
can be solved from Eq.~(\ref{EqBH}).  Since there are more
equations than variables $n(E)$ in general, one can either
discard part of the equations, solving $n(E)$ iteratively, or
use least square method.  In any case, once $n(E)$ is
determined from the set of equations, the canonical average is
computed by
\begin{equation}
 \bigl\langle A(\sigma) \bigr\rangle_T = 
\frac{\displaystyle \sum_E  n(E) 
\bigl\langle A(\sigma) \bigr\rangle_E
 \exp(-E/kT) }{\displaystyle \sum_E n(E) \exp(-E/kT) }. 
\end{equation}

There is no controversy on this part of the theory.  However,
the specific dynamics proposed in ref.~\cite{Oliveira} is
problematic.  The dynamics is a random walk in energy as
follows: pick a site at random, and compute energy change
$\Delta E = E' - E$ if the spin is flipped.  If the new state
has a lower energy, $E'<E$, flip the spin.  If the new state
has a higher energy, $\Delta E>0$, flip the spin with a
probability $\min\bigl(1, N(\sigma, -\Delta E)/N(\sigma,
\Delta E)\bigr)$.  If the energy is the same, $\Delta E = 0$,
flip the spin with some fixed probability, say $1$.

We believe that such a dynamics is wrong in the sense that it
does not give correct microcanonical average with the samples
generated in this dynamics.  The stationary probability
distribution of the dynamics depends not only on energy $E$
but also on the partition $N(\sigma, \Delta E)$ as pointed out
already by Berg and Hansmann \cite{Berg-Hansmann}.  We show
that such a dynamics does not satisfy detailed balance
conditions with respect to an unknown but
energy-dependent-only probability distribution.

Let us first quote a mathematical theorem \cite{Norris} in the
theory of discrete-time Markov chain using notations familiar
to physicists.  As before the symbol $\sigma$ will be a
complete specification of the state (e.g., in Ising model, the
set of all spins).  Let $P_0(\sigma)$ be the probability
distribution of the initial configurations.  If we start from
a unique configuration (for example, from a ground state),
$P_0(\sigma)$ will be 1 for the particular state, and 0 for
all other states.  Let $W(\sigma'|\sigma)$ be the transition
probability, i.e., the probability that the system will be in
state $\sigma'$ given that the current state is $\sigma$;
$W(\sigma'|\sigma) \ge 0$, $\sum_{\sigma'} W(\sigma'|\sigma) =
1$.  Then the probability distribution after $k$ steps of
moves is
\begin{equation}
   P_n(\sigma) = \sum_{\sigma'} W^n(\sigma|\sigma')P_0(\sigma'),  
\end{equation}
where $W^n = W \cdot W \cdot \ldots W$ ($n$ times), viewing
$W$ as a matrix, and $W^n$ is matrix power.  The theorem says
that a unique large-step limit distribution
\begin{equation}
P(\sigma) = \lim_{n\to \infty} P_n(\sigma) 
 = \lim_{n \to \infty}  W^n(\sigma|\sigma')
\end{equation}
exists independent of the initial probability distribution, if
\begin{enumerate}
\item $W$ is irreducible, i.e., $W^k(\sigma'|\sigma) > 0$ for all 
states $\sigma$ and $\sigma'$ and some $k \ge 0$.  In words,
for any initial state $\sigma$ and any final state $\sigma'$,
it is possible with nonzero probability to move from $\sigma$
to $\sigma'$ in $k$ steps, $k$ arbitrary (0, 1, 2, ...) but
finite.  This is usually referred as ergodicity in some
physics literature.

\item $W$ is aperiodic, i.e.,  $W^k(\sigma|\sigma) > 0$
for all $k > k_{min}$, for all $\sigma$.  In words, starting
from state $\sigma$, the probability that it comes back to
state $\sigma$ in $k$ steps is nonzero for all $k$ greater
than a threshold value $k_{min}$.  This point is seldom
mentioned in physics textbooks.  When this condition fails,
one can have probability distributions that oscillate between
two or more forms.  There will be no unique $P(\sigma)$.
\end{enumerate}
Conditions 1 and 2 can be combined into a single statement:
$W^k(\sigma'|\sigma)>0$ for all $\sigma$ and $\sigma'$ and for
all $k$ for sufficiently large $k$.

The theorem guarantees that an equilibrium limit exists and is
unique.  If our aim is to simulate a known distribution, e.g.,
Boltzmann distribution, we also require that $P(\sigma)$ is
that distribution.  It is sufficient then, in addition to 1
and 2, $W$ also satisfies 3---detailed balance,
\begin{equation}
      W(\sigma|\sigma') P(\sigma') = W(\sigma'|\sigma) P(\sigma),
\quad   {\rm for\ all\ } \sigma {\rm\ and\ } \sigma'.
\end{equation}

{\noindent\bf Example 1:} The single-spin-flip Glauber rate
with random selection of sites satisfies 1, 2, and 3 for 
temperature $T>0$ (but not if $T=0$). In this case
\begin{equation}
   W(\sigma'|\sigma) = \cases{0, & \hskip -2.5cm 
   if $\sigma$ and $\sigma'$ are not related by a single spin flip,\cr
    {1 \over 2N}\Bigl(1 - \tanh\frac{\Delta E}{2kT}\Bigr), & 
             $\sigma'$ is obtained by a flip from $\sigma$.\cr}
\end{equation}
The diagonal elements $W(\sigma|\sigma)$ are fixed by
normalization (sum of the column of matrix $W$ is 1, because
probability of going anywhere is 1).

Clearly condition 1 is satisfied---for arbitrary state
$\sigma$ and $\sigma'$, we can move from $\sigma$ to $\sigma'$
by flipping one by one those sites which differ, we can do
this in $k$ steps where $k$ is the number of spins which
differ.  condition 2 is also satisfied.  Clearly
$W(\sigma|\sigma)$ is nonzero for all $\sigma$, so is the
diagonal elements of all power of $W$.  Condition 3 is
satisfied with respect to the Boltzmann equilibrium
distribution.

{\noindent\bf Example 2:} Consider a system of one spin.  In
each step, we flip the spin with probability one.  Then
\begin{equation}
W = \left(\matrix{ 0 & 1 \cr 1 & 0 \cr}\right),
\quad W^{2k} = \left(\matrix{ 1 & 0 \cr 0 & 1 \cr}\right), 
\quad W^{2k+1} = W.
\end{equation}
This stochastic matrix violates condition 2.  Thus, it does not have a
unique limit distribution.

The broad histogram Monte Carlo dynamics in the basic form
(single-spin-flip protocol) is as follows:
\begin{equation}
W(\sigma'|\sigma) = \cases{0, & \hskip -3cm 
                   if $\sigma \neq F \sigma'$ and $\sigma \neq \sigma'$; \cr
   {1 \over N}, & \hskip -3cm 
                        if  $\Delta E \leq 0$ and $\sigma = F\sigma'$; \cr
{1\over N}\min\left(1, {N(\sigma, -\Delta E)\over N(\sigma, \Delta E)}\right), 
                      & if  $\Delta E > 0$ and $\sigma = F\sigma'$. \cr }
\end{equation}
$F$ changes configuration into a new configuration by a
single-spin flip, and $\Delta E = E(\sigma') - E(\sigma)$ is
the energy increment.  $W(\sigma|\sigma)$ is again fixed by
normalization ($ \sum_{\sigma'} W(\sigma'|\sigma) = 1$).  The
factor $(1/N)$ on all of the terms reflects the fact that a
site is picking at random.  $N$ is the total number of spins.

The dynamics has problem with respect to condition 1, as the
ground states are absorbing states.  We cannot go to high
energy states once the system is in the ground state.  This
small problem can be fixed by an ad hoc rule, e.g., if
$N(\sigma, \Delta E) = 0$, $\Delta E < 0$, reset it to
infinity.

Now, we look at condition 3, the detailed balance.  Since we
do not know $P(\sigma)$ for the above dynamics, we assume it
depends on the state only through $E$, i.e., $P(\sigma) =
f\bigl(E(\sigma)\bigr)$ for some unknown function $f(E)$.  
Then consider
two arbitrary states $\sigma$ and $\sigma'$ related by a
single-spin flip with energy $E$ and $E'$, $\Delta E =
E'-E>0$.  From $\sigma$ to $\sigma'$, energy increases,
$W(\sigma'|\sigma) = (1/N)
\min\bigl(1, N(\sigma,-\Delta E)/\allowbreak N(\sigma, \Delta E)\bigr)$.
  From $\sigma'$ to $\sigma$, energy decreases,
$W(\sigma|\sigma') = 1/N$.  Now is
\begin{equation}
    f(E)(1/N) \min\Bigl(1, N(\sigma,-\Delta E)/N(\sigma, \Delta E)\Bigr) 
= f(E') (1/N) 
\end{equation}
for some function $f(E)$?  Clearly, this is not possible in
general unless the ratio $N(\sigma, -\Delta E)/N(\sigma,\Delta
E)$ is a function of $E$ and $E'$ only, and is independent of
the current state $\sigma$.  Within the single-spin-flip
protocol, the ratio does depend on $\sigma$ explicitly.  We
can easily find examples.

$P(\sigma)$ can be computed in principle (and in practice for
small systems) by solving $WP = P$, when the dynamics is
precisely fixed.  Since the configurations in a simulation
appear with probability $P(\sigma)$, the average is ultimately
related to average with the weight of $P(\sigma)$.
Microcanonical average is restored as long as we have
$P(\sigma) = f\bigl(E(\sigma)\bigr)$, i.e., the probability should be
the same for configurations with the same energy.

We present results of a simulation of a 5-spin chain with
periodic boundary condition with the specific dynamics as
discussed above.  The data are presented in Table~1 together
with exact results expected from true microcanonical average.
Monte Carlo steps (MCS) is $10^8$ in the average, so we expect
4 digits of accuracy.

The cases for $E/J = -5$ and 3 agree.  But deviations of order
5\% occur for the case $E/J = -1$.  This is precisely due to
the anomaly in $P(\sigma)$ in the dynamics, since we compute
$\langle N(\sigma, \Delta E)\rangle_E$ by an arithmetic
average:
\begin{eqnarray}
\langle N(\sigma, \Delta E)\rangle_E 
    & = &\mskip -50mu \sum_{\sigma, {\rm\ such\ that\ } E(\sigma)=E} 
 \mskip -40mu
N(\sigma, \Delta E) P(\sigma) /\!\!\!\!\!
  \sum_{\sigma',\; E(\sigma')=E}\!\!\!\! P(\sigma') \\
  & \approx & \sum N(\sigma {\rm \ of\ energy\ } E, \Delta E) 
            / {\rm No.\ of\ samples}. \nonumber
\end{eqnarray}
If $P(\sigma)$ is a constant for a given $E$, then $P(\sigma)$
has no effect.  But if $P(\sigma)$ for $E(\sigma) = E$,
depends on $\sigma$ explicitly, then the above average is not
microcanonical average.

The specific unnormalized $P(\sigma)$s of the five-spin
simulation dynamics are (computed with Mathematica exactly,
which also agree with the simulation), $P(+++++) = 1$,
$P(-++++) = 1$, $P(--+++) = 7/8$, $P(+-+-+) = 5/8$.  Other
degenerate state probabilities are obtained by translation and
spin up-down symmetry.  We also have $N(-++++, 4J) = 2$,
$N(--+++, 4J) = 1$, and $N(-++++, -4J) = 1$, $N(--+++, -4J) =
0$.  From these numbers, we get (averages at $E/J=-1$)
\begin{equation}
\langle N(4J) \rangle  = {2 \cdot 1 + 1 \cdot (7/8)\over 
  1 + 7/8}
              = {23 \over 15} \approx 1.5333,
\end{equation}
\begin{equation}
\langle N(-4J)\rangle  = {1 \cdot 1  + 0 \cdot (7/8) \over 
  1 + 7/8}
              = {8 \over 15} \approx 0.5333,
\end{equation}
which is exactly what we get from Monte Carlo average and it
is wrong!  This feature is generic.  Six and seven spin chains
have the same problem.

In actual implementation, a cumulative average of $N(\sigma,
\Delta E)$ is used instead of the instantaneous value for the
transition rates \cite{Oliveira}.  We have objections to
the procedure.  (1) If one uses the accumulated average $\langle
N(\ldots)\rangle$ up to the current state, then the dynamics
is no longer Markovian---the transition rates not only depend
on the current state but also on all the previous states.
Then the theory of Markov chains does not apply.  
(2) Even if one uses the exact microcanonical average for
the flip rate, the dynamics is still not correct in general.
In Table~2 we present Monte Carlo computations of
$\langle N(\sigma, \Delta E)\rangle$ on a $4 \times 4$
lattice for (a) the original
pure random walk dynamics, (b) the dynamics with the
instantaneous value $N(\sigma, \Delta E)$ replaced by exact
microcanonical average values for the transition rates,
and (c) the true microcanonical average by exact
enumerations.  Clearly, both the original dynamics and
modified one do not give correct answers. 

In a variation of the method \cite{Oliveira-ccp}, the dynamics
is applied to an ensemble with energy limited to small window.
By the same argument above, we see that this is also not
correct.

Detailed balance is a delicate matter.  Since the detailed
balance gives us more equations than the number of
unknowns ($P(\sigma)$) in general,
such equations are inconsistent unless the transition
probabilities have special property.  To illustrate this
point, let us take 
\begin{equation}
W(\sigma'|\sigma) = \delta_{\sigma, F\sigma'} 
r\bigl(E(\sigma')| E(\sigma)\bigr) + \delta_{\sigma, \sigma'}
r\bigl(E(\sigma)|E(\sigma)\bigr),
\end{equation}
 where $\delta_{\sigma, F\sigma'}$ is
1 if $\sigma$ and $\sigma'$ differ by a single spin, and 0
otherwise.  This transition matrix corresponds to a
single-spin-flip protocol with a transition rate $r(E'|E)$
which depends on energies as an arbitrary function.  Does this
transition matrix preserve microcanonical property [$P(\sigma)
= f\bigl(E(\sigma)\bigr)$]?  The answer is not necessarily, unless
\begin{equation}
r(E|E') f(E') = r(E'|E) f(E).  \label{EqEE}
\end{equation}
This can be shown as follows:
\begin{eqnarray}
P_2(\sigma) & = & \sum_{\sigma'} W(\sigma|\sigma')
P(\sigma')\\ 
& = & r\bigl(E(\sigma)|E(\sigma)\bigr)f\bigl(E(\sigma)\bigr)
 + \sum_{E'} N\bigl(\sigma,
E'\!-\!E(\sigma)\bigr) r\bigl(E(\sigma)|E'\bigr) f(E')
\nonumber\\ 
& = & f\bigl(E(\sigma)\bigr) + \nonumber\\ & & \sum_{E'}
N\bigl(\sigma, E'\!-\!E(\sigma)\bigr) 
\Bigl[ r\bigl(E(\sigma)|E'\bigr) f(E') -
r\bigl(E'|E(\sigma)\bigr) f\bigl(E(\sigma)\bigr) \Bigr]. \nonumber
\end{eqnarray}
The last step used the fact that sum of the column elements in
$W$ is one.  The first term depends on $\sigma$ through $E$
implicitly as required.  But there is no guarantee that the
rest of the sum is a function of $E$ only.  However, when
Eq.~(\ref{EqEE}) is satisfied, the rest of the terms are zero.

We can eliminate the unknown function $f(E)$ in
Eq.~(\ref{EqEE}) to get conditions on the matrix elements
themselves.  For example, for three distinct energies $E$,
$E'$, and $E''$ with nonzero transition probabilities among
them, we must have
\cite{Wang,Swendsen},
\begin{equation}
r(E|E'') r(E''|E') r(E'|E) = r(E|E') r(E'|E'') r(E''|E)>0. \label{Eq3T}
\end{equation}
If there are four different energies with nonzero transition
probabilities among them, we should have equations involving
products of four of the transition probabilities.  This would
be the case in three-dimensional Ising model with single-spin
flips.  This equation is not satisfied in the random walk
dynamics with $E$-dependent-only rates where different size of
energy jumps are allowed.

    From the results presented in
refs.~\cite{Oliveira,Oliveira-int,Oliveira-exact,Oliveira-ccp}
the deviations for thermodynamics quantities for large systems
are almost unnoticeable.  Let us discussion in the context of
Eq.~(\ref{Eq3T}) why this is so.  For the two-dimensional Ising
model, this equation is the only relevant constraint equation.
We define
\begin{equation}
 v(E) = \left|1 - 
\frac{r(E|E'') r(E''|E') r(E'|E)}{r(E|E') r(E'|E'') r(E''|E)}\right|
\label{Eqv}
\end{equation}
as the detailed balance violation, where we take $E'=E+4J$,
and $E''=E+8J$.  For the random walk dynamics, 
\begin{equation}
r(E'|E) = \cases{ 1, & if $E'\leq E$; \cr
\min\left( 1, 
\frac{\langle N(\sigma, E-E')\rangle_E}
{\langle N(\sigma, E'-E)\rangle_E}\right). & if $E'>E$.\cr}
\end{equation}
Substituting this expression into Eq.~(\ref{Eqv}), 
we obtain (for $E<0$)
\begin{equation}
 v(E) = \left|1 - 
\frac{ \langle N(\sigma, -4J)\rangle_{E}
       \langle N(\sigma', -4J)\rangle_{E'}   
       \langle N(\sigma, 8J)\rangle_{E}
}{ \langle N(\sigma, 4J)\rangle_{E}
       \langle N(\sigma', 4J)\rangle_{E'}   
       \langle N(\sigma, -8J)\rangle_{E}
}\right|.
\label{Eq19}
\end{equation}
If we take the large-size limit, then the functions $\langle
N(\cdots) \rangle$ are smooth functions in energy, and we
approximate the discrete spectrum by continuous functions,
$n_i(u) = \lim_{N\to\infty} (1/N) \langle N(\sigma, i4J)\rangle_{uN}$, 
$i = 0, \pm 1, \pm 2$, $u=E/N$.
It was pointed out to us by Oliveira \cite{Oliveira-prv} that the
functions $n_i(u)$ can be related to thermodynamics, see also 
ref.~\cite{Moukarzel}.  Here we give a
slightly different argument for it.  Since $n_i(u)$ are
large-size microcanonical average, using the equivalence between
different ensembles in the thermodynamic limit, we can compute the
same qualities by canonical ensemble.  $n_i(u)$ is simply
the probability that a site is surrounded by $i+2$ spins of
the same signs. Evoking Boltzmann distribution, we can show
$n_i(u)/n_{-i}(u) = \exp\bigl(i4J\beta(u)\bigr)$, 
where $\beta(u)=1/kT$, $T$ is the temperature for system
at an average energy $u$ per spin.  Using this result, 
Eq.~(\ref{Eq19}) can be further simplified \cite{Oliveira-prv} netly as 
\begin{equation}
 v(u) = \Bigl|  1 - e^{4J\bigl(\beta(u) - \beta(u+4J/N)\bigr)}  \Bigr|
  \approx - { d \beta \over du }\; {16J^2 \over N}. \label{Eqvres}
\label{Eq20}
\end{equation}

This result suggests that the violation is of order $1/N$ in
general.  At the critical region, it becomes even better, by a
factor of the inverse specific heat.  For the two-dimensional
Ising model at $T_c$ (near $u = u_c = -\sqrt{2}$), we have
$v(u_c) \propto 1/(N \log N)$.  This indeed justifies the
results obtained by computer simulation on large system.  If
instantaneous value $N(\sigma, \Delta E)$ is used, we expect
errors of order $1/\sqrt{N}$.

Since detailed balance violation is a small perturbation to the
transition rate, we expect the Monte Carlo result for $\langle
N(\cdots)\rangle$ also violates detailed balance by an amount
proportional to $v(u)$.  We check this, replacing $r(E'|E)$ by
Monte Carlo estimates of 
$\langle N(\sigma, E'-E)\rangle_E$ in Eq.~(\ref{Eqv}).  Let us
call this quantity $\bar v(u)$.  Numerically we found $\bar
v(u_c) = 0.3, 0.12, 0.03, 0.008$ for two-dimensional square
lattice systems of linear size $L=4$, 8, 16, 32, respectively.
The results are consistent with Eq.~(\ref{Eqvres}).

However, we do not like approximate algorithms.  Finite-size
scalings are important part of computer simulation methods.
It is thus desirable to have exact algorithms with any lattice
sizes.  A remedy to the problem is to use a single function
$f(E)$, similar to ref.~\cite{Lee} (which uses $f(E) \propto
1/n(E)$).  For example, let $f(E) = 1/\langle N(\sigma, \Delta
E)\rangle_E$ for a {\sl fixed\/} $\Delta E$.  Use a flip rate
$r = \min\bigl[1, f(E(\sigma'))/f(E(\sigma))\bigr]$ for a move
from $\sigma$ to $\sigma'$.  This will at least fulfill the
detailed balance.  The second possibility is to restrict
energy changes to only three possible values 0, $\pm 4J$.  The
flip rates can be any arbitrary function of $E$ and $\Delta E
= 0, \pm 4J$.  When such a condition is imposed, we can always
find a consistent solution $f(E)$ to the detailed balance
equations.  This will introduce ergodicity problem, but
inconsistency to equations like (\ref{Eq3T}) will not occur.
This also explains why the dynamics works in one dimension if
$E$-dependent-only flip rates are used.

The last and perhaps also the best recommendation is to take
\begin{equation}
r(E'|E) = \min\left( 1, 
\frac{\langle N(\sigma', E-E')\rangle_{E'}}
{\langle N(\sigma, E'-E)\rangle_{E}}
\right),
\end{equation}
without restriction to energy changes.  When this rate is
used, Eq.~(\ref{Eq3T}) and similar constraint equations
are satisfied automatically (as long as the quantities are
exact microcanonical averages).  This is so because $\langle
N(\sigma, \Delta E)\rangle_E$ themselves are also transitions
rates \cite{Wang} and also satisfy equations similar to
Eq.~(\ref{Eq3T}).  Moreover, we known exactly what is the
stationary distribution $f(E)$.  It is just the inverse
density of states, .i.e., $f(E) \propto 1/n(E)$.  The detailed
balance conditions for this new dynamics are simply equivalent
to Eq.~(\ref{EqBH}).  The dynamics also gives an exactly flat
histogram.  The histogram in general is proportional to $n(E)
f(E)$.  Since $f(E) \propto 1/n(E)$, the energy histogram is a
constant.

For actual implementation, we must determine the
microcanonical average iteratively, using at first cumulative
average of $N(\sigma, \Delta E)$ for transition rates.
After a first approximation, we can either refine them
further, or use the slightly incorrect one but reinforce
Eq.~(\ref{Eq3T}) exactly on data.  When this is done, the
histogram will not be exactly flat, but the dynamics is
correct.  Detail study of this dynamics will be presented
elsewhere.

In conclusion, we have shown that the original random walk
dynamics is only an approximation.  Although it gives good
results for large systems, on small systems, the value for
$\langle N(\cdots)\rangle$ can have errors as large as 20\%.
We proposed new transition rates that satisfy detailed balance
condition.  In particular, instead of using the ratio of
number of moves to states with energy $\pm \Delta E$ from the
current state, we use the ratio of average number of moves from
new states to the current energy states and from current state
to the new states.  This very small change to the original
method plus the requirement that the average values should be
used ensures detailed balance.  In addition, the histogram
obtained by this dynamics is exactly flat.

The author thanks R. H. Swendsen for drawing attention to this
problem. He particularly thanks P. M. C. de Oliveira for 
pointing out errors in the derivations of Eq.~(\ref{Eq19})
and (\ref{Eq20}), for many
clarifications, arguments, stimulating discussions, and
critical readings of several versions of the manuscripts.

\begin{table}
\caption{The average value $\langle N(\sigma, \Delta E)\rangle$
for a system of five-spin chain in broad histogram random walk
dynamics from Monte Carlo simulation and exact results with
true microcanonical average.}
\begin{tabular}{||rcc||}
\hline
   $E/J$ &  $\langle N(\sigma, \Delta E=-4J) \rangle_E$ &  
$\langle  N(\sigma, \Delta E=+4J)\rangle_E$ \\
\hline
\hline   
\multicolumn{3}{|c|}{Random Walk Dynamics Result} \\
\hline
    $-5$    &    0        &      5   \\
    $-1$    & 0.53326     & 1.53326  \\
      3     &    3        &      0   \\
\hline
\hline    
\multicolumn{3}{|c|}{Exact Microcanonical Results} \\
\hline    
   $-5$   &    0     &     5  \\
   $-1$   &   1/2    &    3/2 \\
     3    &    3     &     0  \\
\hline
\hline
\end{tabular}
\medskip
\caption{Monte Carlo random walk dynamics and exact results
on a $4 \times 4$ square lattice with periodic boundary
conditions: (a) pure broad histogram random walk dynamics,
using instantaneous $N(\sigma, \Delta E)$, ${\rm
MCS}=1.45\times 10^9$; (b) broad histogram random walk
dynamics with strictly $E$-dependent flip rates, using the
exact $\langle N(\sigma,\Delta E) \rangle_E$, ${\rm
MCS}=1.45\times 10^{10}$; (c) true microcanonical average
computed by exact numeration of all the states.
Statistical/numerical error is on the last digit.}
\begin{tabular}{||clll||}
\hline
$E/(4J)$ &  
$\langle N(\sigma,\!\Delta\! E\!=\!\!-\!8J) \rangle_E$ &  
$\langle N(\sigma,\!\Delta\! E\!=\!\!-\!4J) \rangle_E$ &
$\langle N(\sigma,\!\Delta\! E\!=\!0) \rangle_E$ \\
\hline
\hline   
\multicolumn{4}{|c|}{(a) Random walk dynamics, 
current $N(\sigma, \Delta E)$} \\
\hline
 4 &  0.9878 &  0.7850 &  1.6194 \\
 5 &  0.4348 &  1.5815 &  3.0233 \\
 6 &  0.4077 &  1.7295 &  5.4252 \\
 7 &  0.5626 &  2.6570 &  5.9934 \\
\hline
\hline    
\multicolumn{4}{|c|}{(b) Random walk using 
average $\langle N(\sigma, \Delta E)\rangle$} \\
\hline    
 4 &  0.86573  & 0.80220  & 1.6888   \\
 5 &  0.39313  & 1.5878   & 3.2620   \\
 6 &  0.47274  & 1.7930   & 5.3393 \\
 7 &  0.48124  & 2.9319   & 5.8429 \\
\hline
\hline    
\multicolumn{4}{|c|}{(c) Exact microcanonical average} \\
\hline    
 4 & 0.83018868 & 0.90566038 & 1.81132075 \\
 5 & 0.29629630 & 1.55555556 & 3.55555556 \\
 6 & 0.38755981 & 1.81818182 & 5.51196172 \\
 7 & 0.45283019 & 2.94339623 & 5.88679245 \\
\hline
\hline    
\end{tabular}
\end{table}

\end{document}